\documentclass[a4paper,12pt]{article}
\usepackage{amsfonts,amsmath}

\begin{document}
\title{\textbf{Surface waves in orthotropic 
incompressible materials}}

\author{Michel Destrade}
\date{2001}
\maketitle
%
\begin{abstract}
{\small
The secular equation for surface acoustic waves propagating 
on an orthotropic incompressible half-space is  derived in a direct 
manner, using the method of first integrals.\\
}
\end{abstract}

\newpage


\section{INTRODUCTION}
{\normalsize
The problem of elastic waves propagating on the free surface of a
semi-infinite elastic body is a well-covered research topic, 
initiated by
Rayleigh \cite{Rayl85} in his study of seismic waves within the context
of classical linear  elasticity.
For anisotropic crystals, Barnett and Lothe \cite{BaLo85} have
drawn on the works of Stroh \cite{Stro62} to build a complete theory
of surface waves based on an ana\-lo\-gy between surface wave
propagation and straight line dislocation motion.
Extensive coverage and surveys of that topic can be found, for
instance, in a textbook by Ting \cite{Ting96}.

Recently, there has been some interest \cite{SoNa99,NaSo99,NaSo97} in
the study of wave propagation in ani\-so\-tro\-pic materials subjected to
the constraint of \textit{incompressibility}.
The purpose of the present paper is to establish the secular 
equation for surface (Rayleigh) waves propagating on the free 
plane surface of an incompressible orthotropic half-space.
A similar problem was solved by Chadwick \cite{Chad97} within the 
context of finite elasticity: 
he considered the propagation of small-amplitude surface waves in a 
finitely deformed incompressible material;
the deformation was static and purely homogeneous, and the strain 
energy function for the incompressible nonlinearly elastic 
material was such that the 
deformed body presented orthotropic anisotropy.
Following Nair and Sotiropoulos \cite{NaSo97}, the present article  
focuses on 
an orthotropic linearly elastic material for which the usual 
stress-strain relations are modified to take the incompressibility
constraint into account, by adding an isotropic pressure term.
These authors have argued that 
``the assumptions of incompressibility and orthotropy are applicable 
to several materials such as, for example, polymer Kratons, 
thermoplastic elastomers, rubber composites when low frequency waves 
are considered to justify the assumption of material homogeneity, 
etc.''
Other studies use these assumptions for the modeling of laminated
composites made alternatively with reinforcing (filler) layers and
matrix (binder) layers \cite{GuGu99}, or with stiff fibers and 
incompressible epoxy matrices \cite{Sutc92}.

The primary purpose of this paper is to show that the method of first
integrals used by Mozhaev \cite{Mozh94} to derive, in a rapid and 
elegant manner, the secular equation for surface waves in 
(compressible) orthotropic materials, can also be employed in the 
case of incompressible orthotropic materials.
This can be achieved by applying the method of first integrals to a
system of second order ordinary differential equations for the
components of the \textit{tractions} on surfaces parallel 
to the free surface, 
rather than for the components of the mechanical displacement 
(as in Ref. \cite{Mozh94}). 
In the latter case, the pressure appears in the system of differential
equations, whereas in the former case, it does not, and hence the
number of unknowns is reduced from four 
(the pressure and the components of the mechanical displacement) 
to three (the components  of the traction on  surfaces parallel 
to the free surface).
Also, the mechanical boundary conditions are easily written, because
they correspond to the nullity of these traction components on the 
free surface of the half-space, and at infinite distance 
from this surface.
A third advantage of this approach is that the assumption of plane
strain \cite{NaSo97} is not required a priori.

The paper is  organized as follows.
In Section II,  the basic equations governing the propagation
of elastic waves in an orthotropic incompressible material
are recalled.
In Section III,  these equations are written for the case of surface 
acoustic waves.
Then  a system of six first order differential equations 
for the displacement and the traction components is derived. 
Eventually a system of three second order differential equations 
is found for the traction components. 
One of these three equations is trivially solved when the 
boundary conditions are applied.
In Section IV, the method of first integrals \cite{Yu83, Mozh94} 
is applied to the two remaining equations, and 
the secular equation for surface waves in orthotropic incompressible 
materials is  quickly derived.
As a check,  the isotropic case is treated and Rayleigh's 
original equation \cite{Rayl85} is recovered.
Also, the correspondence between this paper's result and Chadwick's 
result \cite{Chad97} is shown. 
Finally in Section V, possible developments for this work are 
presented.

\section{PRELIMINARIES}

First,  the governing equations for an incompressible
orthotropic elastic material are recalled.
The material axes of the body are denoted
by $x_1$, $x_2$, and $x_3$.
The equations may be derived from the classical linearized equations 
of anisotropic elasticity \cite{Love27} by adding an isotropic pressure
term $p \mathbf{1}$ (say) to the nominal stress 
$\mbox{\boldmath $\sigma$}$ 
(say).
Hence, for orthotropic incompressible elastic bodies \cite{NaSo97},
\begin{equation} \label{StressStrainGeneral}
\begin{array}{l}
\sigma_{11}=-p 
 + C_{11} \epsilon_{11} +C_{12} \epsilon_{22}+C_{13} \epsilon_{33},
\\
\sigma_{22}=-p 
 + C_{12} \epsilon_{11} +C_{22} \epsilon_{22}+C_{23} \epsilon_{33},
\\
\sigma_{33}=-p 
 + C_{13} \epsilon_{11} +C_{23} \epsilon_{22}+C_{33} \epsilon_{33},
\\
\sigma_{32}= 2 C_{44} \epsilon_{32}, \quad  
\sigma_{13}=2 C_{55} \epsilon_{31}, \quad 
\sigma_{12}= 2 C_{66} \epsilon_{12}, 
\end{array}
\end{equation}
where $\epsilon$'s denote the strain components, and $C$'s the 
elastic constants. 
The strain components are related to the 
displacement components $u_1$, $u_2$, $u_3$ through
\begin{equation}
\epsilon_{ij}=(u_{i,j}+u_{j,i})/2 \quad (i,j=1,2,3).
\end{equation}
Finally, the incompressibility constraint reads
\begin{equation} \label{IncomprGeneral}
u_{1,1}+u_{2,2}+u_{3,3}=0,
\end{equation}
and the equations of motion, in the absence of body forces,
are written as
\begin{equation} \label{EqnMotnGeneral}
\sigma_{ij,j}= \rho u_{i,tt} \quad (i=1,2,3),
\end{equation}
where $\rho$ is the mass density of the material, 
and the comma denotes differentiation.
These are the equations established by Nair and Sotiropoulos 
\cite{NaSo97}. 
These authors also note that for plane strain deformations, 
the strain-energy function density is positive
definite when the following inequalities are satisfied,
\begin{equation}
C_{66} \ge 0, \quad C_{11}+C_{22}-2C_{12} \ge 0.
\end{equation}

\section{SURFACE WAVES}

Here the equations of motion for a surface wave in a 
semi-infinite body made of an orthotropic incompressible elastic 
material are established.
Attention is restricted to propagating 
inhomogeneous surface waves which are
subsonic with respect to homogeneous body waves.
The modelisation of the surface wave follows that of Mozhaev 
\cite{Mozh94}:
the plane wave propagates with speed $v$, wave number $k$, 
and corresponding displacement and pressure of the form 
\begin{equation}
[u_j(x_1,x_2,x_3), p(x_1,x_2,x_3)]
=[U_j(x_2),k P(x_2)] e^{ik(x_1 -vt)} \quad (j=1,2,3),
\end{equation}
where the $U$'s and $P$ are unknowns functions of $x_2$ alone.
For these waves, the planes of constant phase are orthogonal to the
$x_1$-axis, and the planes of constant amplitude are orthogonal to the
$x_2$-axis.
The stress-strain relations \eqref{StressStrainGeneral} reduce to 
\begin{equation}\label{StressStrain}
\begin{array}{l}
t_{11}=-P  +i C_{11}U_1 +C_{12}U_2',
\\
t_{22}=-P  +i C_{12}U_1 +C_{22}U_2',
\\
t_{33}=-P  +i C_{13}U_1 +C_{23}U_2',
\\
t_{32}=C_{44} U_3', \quad  
t_{13}= iC_{55} U_3, \quad 
t_{12}= C_{66} (U_1'+iU_2), 
\end{array}
\end{equation}
where the prime denotes differentiation with respect to $k x_2$,
and the $t$'s are defined by 
\begin{equation} 
\sigma_{ij}(x_1,x_2,x_3)= k t_{ij}(x_2)e^{ik(x_1 -vt)}
\quad (i,j=1,2,3).
\end{equation}

The surface $x_2=0$ is assumed to be free of tractions, and 
the mechanical displacement and pressure are assumed to be 
vanishing as $x_2$ tends to infinity. 
These conditions lead to the following boundary conditions,
\begin{equation} \label{BC1}
t_{i2}(0)=0, \quad U_i(\infty)=0 
\quad (i=1,2,3),\quad P(\infty)=0.
\end{equation}

Finally, the equations of motion \eqref{EqnMotnGeneral} and the 
incompressibility constraint \eqref{IncomprGeneral} reduce to 
\begin{equation} \label{EqnMotn}
\begin{array}{l}
i t_{11}+ t_{12}'= - \rho v^2 U_1, \:
i t_{12}+ t_{22}'= - \rho v^2 U_2, \:
i t_{13}+ t_{32}'= - \rho v^2 U_3, 
\\
i U_1 + U_2' =0.
\end{array}
\end{equation}

Note that a classical approach would be to substitute in this 
last equations, the expressions obtained earlier for the stress
tensor components, which would lead to a system of four second order
differential equations for the unknown functions $U_1, U_2, U_3, P$.
Instead,  the Stroh formalism is now used 
to derive a system of six first
order differential equations for the components of the displacement
and the tractions on the surface $x_2=$const.
Thus, introducing the notation
\begin{equation}
t_i = t_{i2} \quad (i=1,2,3),
\end{equation} 
and using Eqs.~\eqref{StressStrain}-\eqref{EqnMotn}, 
the system is found as
\begin{equation} \label{Stroh}
\begin{array}{l}
U_1'= -i U_2 + (1/C_{66}) t_1 , \quad
U_2'= -i U_1, \quad
U_3'= (1/C_{44}) t_3 , \quad
\\
t_1'=(C_{11}+C_{22}-2C_{12}-\rho v^2) U_1 - i t_2,   \:
t_2'= -\rho v^2 U_2 - i t_1, \:
t_3'=(C_{55}-\rho v^2) U_3.
\end{array}
\end{equation}

Now a system of three second order differential equations for 
$t_1,t_2,t_3$  is derived as follows.
First, differentiation of \eqref{Stroh}$_{4-6}$ yields relations
between the $t_i''$ and the $u_i', t_i'$, or equivalently, using
\eqref{Stroh}$_{1-3}$ between the $t''_i$ and the $u_i, t_i', t_i$.
Then, substitution for the $u_i$ by their expression in terms of 
the $t_i', t_i$ obtained from \eqref{Stroh}$_{4-6}$ is performed.
Eventually it is found that the  $t_i'', t_i', t_i$ $(i=1,2,3)$ 
must satisfy the following equations,
\begin{equation} \label{system}
\begin{array}{l}
(\rho v^2) t_1''
 - i(C_{11}+C_{22}-2C_{12}-2\rho v^2)t_2' \\
  \qquad 
  +(C_{11}+C_{22}-2C_{12}-\rho v^2)(1-\rho v^2/C_{66})t_1
   =0, \\
(C_{11}+C_{22}-2C_{12}-\rho v^2)t_2'' 
  +i(C_{11}+C_{22}-2C_{12}-2\rho v^2)t_1'
   + \rho v^2 t_2
     =0, \\
C_{44} t_3'' - (C_{55} - \rho v^2) t_3 =0,
\end{array}
\end{equation}
and are subject to the following boundary conditions,
\begin{equation} \label{BC2}
t_{i}(0)=  t_i(\infty)= 0 
\quad (i=1,2,3).
\end{equation}

The third differential equation in the system \eqref{system} 
is decoupled from the two others, and  can be solved exactly.
Taking the boundary conditions \eqref{BC2}$_3$ into account, 
it is seen that
\begin{equation}
t_3(x_2)=0, \quad \text{for all } x_2,
\end{equation}
and hence the motion is a pure mode \cite{Chad76} for the tractions
on the surface $x_2=$ const.
Now  the coupled system of the two remaining equations may be solved.

\section{SECULAR EQUATION}

For surface waves in compressible orthotropic materials, Mozhaev
\cite{Mozh94} applied the method of first integrals to a system of 
two differential equations for the two nonzero components of the
mechanical displacement.
Here  a similar procedure for the two nonzero components
$t_1, t_2$ of the tractions on the surface $x_2=$const. is followed,
and  the secular equation for surface waves in incompressible 
orthotropic materials is obtained in a direct manner.

The differential equations \eqref{system}$_{1,2}$ 
for $t_1, t_2$ are expressed as
\begin{equation} \label{system2}
\begin{array}{l}
\xi t_1'' +i(\delta -2 \xi)t_2' -(\delta - \xi)(1-\xi) t_1 =0, \\
(\delta -\xi) t_2'' -i(\delta -2 \xi)t_1' -\xi t_2 =0, 
\end{array}
\end{equation}
where $\xi$ and $\delta$ are defined by
\begin{equation}
\xi=(\rho v^2)/C_{66}, \quad \delta=(C_{11}+C_{22}-2C_{12})/C_{66}.
\end{equation}

The speed given by $\xi =1$ (that is, $\rho v^2 = C_{66}$) corresponds
to the speed of a body (homogeneous) wave propagating in the 
$x_1$-direction, and gives therefore an upper bound for the speed of 
subsonic surface waves.
Throughout the rest of paper, it is assumed that the surface wave 
travels with a speed distinct from that given by $\xi = \delta$ 
(that is, $\rho v^2 \ne  (C_{11}+C_{22}-2C_{12})/C_{66}$).

Now  multiplication of  \eqref{system2}$_1$ by $t_1'$ 
and \eqref{system2}$_2$ by $t_2'$, 
and integration  between $x_2=0$ and $x_2=\infty$, yields,
using the boundary conditions \eqref{BC2},
\begin{equation}
\xi t_1'(0)^2 -2i(\delta-2\xi) \textstyle{\int} t_1' t_2' =0, 
\quad \text{and} \quad
(\delta - \xi)t_2'(0)^2 +2i(\delta-2\xi) 
\textstyle{\int} t_1' t_2' =0, 
\end{equation}
so that 
\begin{equation} \label{eq1}
\xi t_1'(0)^2 + (\delta - \xi)t_2'(0)^2 =0.
\end{equation}
Similarly, multiplication of \eqref{system2}$_1$ by 
$\xi t_1'+i(\delta -2 \xi)t_2$ and \eqref{system2}$_2$ 
by $(\delta -\xi) t_2' -i(\delta -2 \xi)t_1 $, 
and integration between $x_2=0$ and $x_2=\infty$, yields
\begin{multline}
\xi^2 t_1'(0)^2 +2i(\delta-2\xi)(\delta-\xi)(1-\xi) 
\textstyle{\int} t_1 t_2 =0, 
\\ \quad \text{and} \quad
(\delta - \xi)^2 t_2'(0)^2 -2i(\delta-2\xi)\xi 
\textstyle{\int} t_1 t_2 =0, 
\end{multline}
so that 
\begin{equation} \label{eq2}
\xi^3  t_1'(0)^2 + (\delta - \xi)^3 (1-\xi)t_2'(0)^2 =0.
\end{equation}
Eqs.\eqref{eq1} and \eqref{eq2} form a trivial 
system of two equations for the unknowns $t_1'(0)^2$ and $t_2'(0)^2$, 
whose determinant must be zero:
\begin{equation}
\xi (\delta - \xi)[(\delta - \xi)^2 (1-\xi)- \xi^2]=0.
\end{equation}
It follows that the \textit{secular equation} is given by
\begin{equation} \label{secular}
(\delta - \xi)^2 (1-\xi)= \xi^2, \quad \text{i.e.} \quad
(C_{11}+C_{22} -2C_{12}-\rho v^2)^2 (C_{66}-\rho v^2)
= C_{66} (\rho v^2)^2.
\end{equation}
This equation constitutes the  main result of the paper: 
the direct and explicit derivation of the secular equation 
for subsonic surface waves propagating in a semi-infinite body made 
of orthotropic incompressible linearly elastic material.
It is worth mentioning that this result can be used for other types
of anisotropy: Royer and Dieulesaint \cite{RoDi84} have indeed 
proved that with respect to surface waves, 
results established for the orthotropic case may be
applied to 16 different configurations, including cubic, tetragonal,
and hexagonal anisotropy.

In order to justify the existence of a real wave speed,
the secular equation \eqref{secular} is expressed as 
\begin{equation}
f(\xi)=0, \quad \text{where} \quad
f(\xi)= \xi^2-(\delta - \xi)^2 (1-\xi).
\end{equation}
As noted earlier, for traveling subsonic surface waves, 
this secular equation is subject to 
\begin{equation} \label{range}
0 \le \xi \le 1.
\end{equation}
Within this range, it is easy to prove that $f$ is a 
monotonic increasing  function of $\xi$, and that
\begin{equation}
f(0)= - \delta^2, \quad f(1) = 1.
\end{equation}
It follows that the secular equation has a unique positive root 
in the interval \eqref{range}.

For consistency purposes, the main result established in this paper is
related to previous studies.
First,  attention is given to the isotropic limit, when 
$C_{11}=C_{22}=\lambda+2\mu$, $C_{12}=\lambda$, $C_{66}=\mu$,
where $\lambda$ and $\mu$ are the classical Lam\'e moduli of 
elasticity.
In this case,  the secular equation, written for $\xi=\rho v^2/ \mu$,
reduces to 
\begin{equation}
(4 - \xi)^2 (1-\xi)= \xi^2, \quad \text{or} \quad
\xi^3 -8\xi^2 +24\xi -16=0,
\end{equation}
which is the well-known equation derived by Lord Rayleigh
\cite{Rayl85}, by considering the incompressible limit 
($\lambda=\infty$) for an isotropic linear elastic material.

Next, another previous result  is put into perspective.
Chadwick \cite{Chad97} has  adapted the Stroh 
formalism to the theory of prestressed incompressible nonlinearly 
elastic materials.
Considering a material whose stored energy function is such that the
body will present orthorhombic anisotropy once it has been subjected
to a large pure homogeneous deformation, 
he obtained the secular equation for 
surface waves propagating in a principal direction as
\begin{equation} \label{Chadwick}
[2(B+C- \overline{\sigma}) - \rho v^2][C(A - \rho v^2)]^{1/2}
=(C- \overline{\sigma})^2 -C(A - \rho v^2),
\end{equation}
where $A,B,C$ are constants defined in terms of the strain energy,
initial pressure, and initial stretch ratios,
and $\overline{\sigma}$ is the normal stress applied on the surface
$x_2=0$.
When this surface is free of tractions,  $\overline{\sigma}=0$ and
after squaring, Eq.~\eqref{Chadwick} reduces to
\begin{equation}
(2B+C-A-\eta^2)^2 (C-\eta^2)=C (\eta^2)^2,
\end{equation}
where $\eta^2= C-A+ \rho v^2$.
This equation may be formally compared to Eq.~\eqref{secular}$_2$,
where $\eta^2$, $C$, and $2B-A$ play the role of $\rho v^2$, 
$C_{66}$, and $C_{11}+C_{22}-C_{66} -2C_{12}$, respectively.

Finally, Nair and Sotiropoulos \cite{NaSo99} 
have obtained an \textit{implicit} form of the secular equation
for surface waves propagating in a monoclinic incompressible material.
By taking the elastic coefficients $C_{16}$ and $C_{26}$ to be zero
in their analysis, the reader may check that the explicit secular
equation \eqref{secular} is recovered.

\section{DISCUSSION}

The secular equation for surface waves
on an incompressible orthotropic half-space was derived directly. 
Hence it has been shown that a powerful method presented by 
Mozhaev \cite{Mozh94}, but which seems 
to have remained unnoticed, can be adapted to take the 
constraint of incompressibility into account. 

For monoclinic or triclinic materials, 
the method of first integrals cannot be applied in the case of a 
three dimensional displacement.
As demonstrated by Mozhaev \cite{Mozh94}, 
it leads to a trivial system of 18 equations for 18 unknowns, 
but the rank of the system turns out to
be 17 at most, a fact which appears to have been overlooked by the 
author.

However, for \textit{plane strain} deformations, 
some further results may be established.
For instance, Sotiropoulos and Nair \cite{SoNa99} have studied the 
reflection of plane elastic waves from a free surface in 
incompressible monoclinic materials with plane of symmetry at $x_3=0$, 
and Nair and Sotiropoulos \cite{NaSo99} have considered interfacial
waves with an interlayer in the same type of materials.
In particular, they derived the secular equation for surface
(Rayleigh) waves in an implicit form.
The first integrals method makes it possible to write the secular 
equation in explicit form, as is proved in a forthcoming article.
Possibly, interfacial (Stoneley) waves may also be investigated.


\end{document}